\documentstyle[aps, epsf, 12pt, axodraw]{revtex}
\tightenlines
\begin{document}

\newcommand{\be}{\begin{equation}}
\newcommand{\ee}{\end{equation}}
\newcommand{\bi}{\bibitem}
\newcommand{\bea}{\begin{eqnarray}}
\newcommand{\eea}{\end{eqnarray}}
\newcommand{\nn}{\nonumber}
\newcommand{\ps}{p\kern-0.175cm /}
\newcommand{\psm}{\left(\ps + m \right)}
\newcommand{\ks}{k\kern-0.19cm /}
\newcommand{\prk}{\frac{i}{\ps-\ks-m}}
\newcommand{\prkd}{\frac{i}{-2 p k  + k^2}}
\newcommand{\prkn}{\left( \ps - \ks + m \right)}
\newcommand{\prko}{\frac{i}{\ps-\ks_1-m}}
\newcommand{\prkon}{\left( \ps - \ks_1 + m \right)}
\newcommand{\prkod}{\frac{i}{-2 p k_1  +k_1^2}}
\renewcommand{\prc}{\frac{i}{\ps-\ks_c-m}}
\newcommand{\prbc}{\frac{i}{\ps-\ks_b-\ks_c-m}}
\newcommand{\prdbc}{\frac{i}{-2 p \left(k_b +k_c \right) + \left( k_b + k_c
\right)^2}}
\newcommand{\prdc}{\frac{i}{-2 p k_c + k_c^2}}
\newcommand{\prkko}{\frac{i}{\ps-\ks-\ks_1-m}}
\newcommand{\prkkon}{\left( \ps - \ks- \ks_1 + m \right)}
\newcommand{\prkkod}{\frac{i}{-2 p \left(k + k_1 \right) + 
\left( k + k_1 \right)^2}}
\newcommand{\gk}{\frac{-i}{k^2}}
\newcommand{\gko}{\frac{-i}{k_1^2}}
\newcommand{\gkc}{\frac{-i}{k_c^2}}
\newcommand{\gkb}{\frac{-i}{k_b^2}}
\newcommand{\gkbc}{\frac{-i}{(k_b+k_c)^2}}
\newcommand{\ph}{d^4 k d^4 k_1}
\newcommand{\phbc}{d^4 k_b d^4 k_c}

\newcommand{\gua}{\gamma^{\alpha}}
\newcommand{\guc}{\gamma^{\delta}}
\newcommand{\guai}{i g \gamma^{\alpha}}
\newcommand{\guat}{i g \gamma^{\alpha} T^a}
\newcommand{\guct}{i g \gamma^{\delta} T^c}
\newcommand{\gub}{\gamma^{\beta}}
\newcommand{\gubi}{i g \gamma^{\beta}}
\newcommand{\gubt}{i g \gamma^{\beta} T^b}
\newcommand{\gdai}{i g \gamma_{\alpha}}
\newcommand{\gda}{\gamma_{\alpha}}
\newcommand{\gdat}{i g \gamma_{\alpha} T^a}
\newcommand{\gdb}{\gamma_{\beta}}
\newcommand{\gdbi}{i g \gamma_{\beta}}
\newcommand{\gdbt}{i g \gamma_{\beta} T^b}
\newcommand{\go}{\gamma^{0}}

\newcommand{\sv}{S^{abc}_{\alpha \beta \delta}}
\newcommand{\cf}{C_F^2}
\newcommand{\ca}{\frac{C_F C_A}{2}}
\newcommand{\cfa}{\left(C_F^2-\frac{C_F C_A}{2}\right)}

\newcommand{\ub}{\bar{u}}
\newcommand{\mb}{\bar{m}}
\newcommand{\vea}{\left[vertex\right]_{\alpha}^a}
\newcommand{\veb}{\left[vertex\right]_{\beta}^b}

\newcommand{\fif}{C 5im_b^4}
\newcommand{\tif}{C 10m_b^3}
\newcommand{\zet}{\left( Z^{-1} - 1 \right)}

\title{Cancellation of $1/m_Q$ Corrections to the Inclusive Decay Width of a
Heavy Quark 
\footnote{Talk given at the 36th International School of 
Subnuclear Physics, Erice}
}
\author{A. Sinkovics}
\address{
The Randall Laboratory of Physics\\
University of Michigan\\
Ann Arbor, MI 48109-1120}
\maketitle


\section{Introduction} \label{sec:intro}

This talk is a short summary of a work done in collaboration with 
Ratindranath Akhoury and Valentine Zakharov \cite{paper}, a study of power
corrections to the inclusive semileptonic decay of a heavy quark. 
The phenomenological motivation of this work is that the semileptonic
inclusive decay can provide information on the CKM matrix elements.
Power corrections may be large and knowledge about them is important 
for the precision extraction of the CKM parameters.

Inclusive decays of hadrons containing a single heavy quark (mass $m_b$) 
have been studied \cite{hqope} within the operator product expansion (OPE) 
which gives an expansion of the decay width in terms of a small parameter 
$\Lambda/m_b$ :
\be
\Gamma_{incl}~=~\Gamma^{pert}_{incl}(1+ a_1 \Lambda/m_b + a_2 ( \Lambda/m_b)^2
+...).
\label{rate}
\ee
In the above, $\Gamma^{pert}_{incl}$ is the decay width of the heavy quark
including all perturbative corrections (the parton model result). 
Since there are no local gauge invariant operators of the appropriate 
dimension, an important conclusion \cite{hqope} is that $a_1=0$, and the higher order power corrections can be classified in terms of the matrix elements of various operators.

Power corrections may be studied with the method of renormalons \cite{renorm}
within perturbation theory.
The basic idea of the renormalon method is
that one can obtain information about power corrections by looking at a class of
diagrams which give factorial divergence in large orders of perturbation theory.
The power corrections are seen to arise from regions of low momenta. This leads
 us to investigate the infrared sensitivity of inclusive decays, and we can
replace the QCD scale $\Lambda$, which is a non-perturbative parameter, 
by an infrared cutoff parameter $\lambda$ defined within perturbation theory. 
In the case of QED, for example, it could be a (fictitious) photon
mass. Then the inclusive rate (\ref{rate}) may be represented as :

\be
\Gamma_{incl}~=~\Gamma^{pert}_{incl}(1+ b_1 \lambda/m_b + b_2 ( \lambda/m_b)^2 ln
\lambda  +...). \label{prate}
\ee
Note that the Landau conditions for the singularities of Feynman diagrams tell
us that the infrared sensitive contributions arise only as terms which are 
non-analytic in $\lambda^2$. 

An advantage of the OPE is the generality of its applicability to 
non-perturbative contributions. On the other hand, the conditions for
the validity of the OPE can be checked by explicit perturbative
calculations. As mentioned before, since no dimension 4 local gauge invariant
operator in HQET, the coefficient of the first, linear correction $b_1=0$.
Explicit one-loop order calculations \cite{lincan} 
however have found linear correction to the decay width ($b_1\neq 0$).
The reason is that the pole mass is infrared sensitive, 
so on-shell renormalization must be avoided, and one has to use a 
short-distance mass instead. The decay width has no linear correction only if
expressed not in terms of the pole mass $m_b$ but in
terms of a short distance or a $\bar{MS}$ mass. 
In this way one avoids an otherwise large (possibly
non-universal) correction of order $1/m_b$, but at the expense of
introducing a renormalization scale ($\mu$) dependence in the total rate through
the use of the running mass. 
 
We extended the above argument for the inclusive 
semileptonic decay of a heavy quark in QCD to the second loop order when the 
non-abelian nature of the interactions are first apparent. 
That is we show the cancellations of all terms linear in the infrared
cutoff in the decay width when the latter is expressed in terms of the short
distance mass. The possible sources of infrared sensitive terms
are the mass and wave function renormalization diagrams of the initial 
heavy quark and the bremsstrahlung diagrams. 
The cancellations proceed in a different manner for the two
cases because whereas the wave function renormalization is multiplicative, the
on-mass shell renormalization is additive. 


\section{Connection to the Kinoshita-Lee-Nauenberg theorem}

Certain questions immediately come to mind concerning the generality of the
result. Is there a general principle behind this cancellation ?, and is it true
to all orders in perturbation theory, even for the non-abelian case? In fact
there are infrared safe observables in QCD that do receive power corrections
that are proportional to $1/M$ where $M$ is a large mass scale. Examples are
provided by the event shape variables in $e^+e^-$ annihilation \cite{thrust} for
which there does not exist an operator product expansion. It has been argued in
ref. \cite{az1} that inclusive enough observables do not receive $1/M$ power
corrections whereas the more exclusice ones that assume some precision 
measurement do.
Below we will apply this principle to the inclusive decay of the heavy quark, 
and argue that the KLN \cite{kln} theorem is behind the cancellation of the 
terms proportional to $1/m_b$. 

The KLN theorem states the cancellation of infrared divergences in the
transition probability when it is summed over an appropriate degenerate set of initial and final states.
In a recent publication \cite{asz} it was argued that in the KLN sum
not only the leading infrared sensitive piece
proportional to $ln \lambda$ but also the next, subleading 
$\sim \lambda$ term is cancelled, i.e,
\be
\sum_{i,f} |S_{i\rightarrow f}|^2 \sim 0 \cdot ln \lambda + 0 \cdot 
\lambda + {\rm terms~independent~of~ \lambda} + O(\lambda^2 ln \lambda).
\ee
The transition probability summed over both initial and final states cannot be
related to a physical inclusive cross section. However we argue that
for heavy quark decay the initial state is trivial and hence from the above
the KLN theorem guarantees the cancellation not only of
the $\ln \lambda$ terms but also the subleading linear pieces.
According to the uncertainity principle, the total energy can be measured to
any accuracy provided the measurement time is long enough.
In particular, the uncertainity $\Delta
E$ in the total energy can be made smaller than an infrared cutoff:
\be
\Delta E \ll \lambda
\ee
Now consider a decaying charged (or colored) particle in its rest frame. It is
obvious that this state is not degenerate with any other since there is a
gap between the mass of the charged particle and the energies in the continuum
for any $\lambda \neq 0$. 
Thus 
the summation over the initial states in the KLN theorem is redundant
for this case. 
Hence we conclude that
\be
\Gamma_{incl}~=~\Gamma^{pert}_{incl}\left(1 + O(\lambda^2 ln \lambda)\right).
\ee
It should be emphasized that this argument guarantees the above equality to all
orders in perturbation theory. However it is very important
for the KLN theorem to be valid that the renormalization procedure does not
introduce any infrared sensitivity. Thus since the pole mass is infrared
sensitive, on shell mass renormalization must be avoided in favour of the
$\bar{MS}$ scheme . We would also like to point out that the same
argument can be applied not just to the semileptonic decay but to other inclusive
decays like for example the radiative one, $B \rightarrow X_s \gamma$. 

As mentioned before, we verify the above argument by an explicit second loop 
order calculation: we show the cancellation of terms linearly proportional to
an infrared cutoff $\lambda$. Our method, 
dicussed in section \ref{sec:prelim},
is to introduce local effective vertices for the diagrams with radiation from 
the final state quark to linear accuracy.
 
\section{Effective vertices} \label{sec:prelim}

We consider the totally inclusive semileptonic decay 
of a heavy quark and examine the possible terms linearly proportional to an
infrared cutoff. Such terms can arise from the soft gluon contributions to the
diagrams for the mass and wave function renormalizations of the heavy quark and
from the bremsstrahlung diagrams.  

\begin{figure}[p]
%
%
%

\def\epsfsize#1#2{0.75#1}
\centerline{\epsfbox{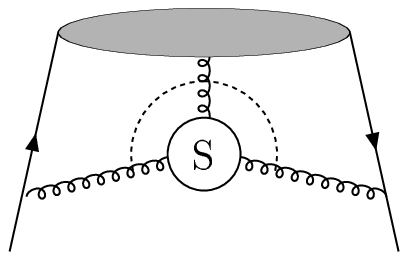}}
\bigskip

\caption{\setlength{\baselineskip}{0.30in} 
\setlength{\baselineskip}{0.30in}
The forward scattering amlitude for the heavy quark decay whose 
imaginary part gives the decay width. Shaded blob denotes
the hard part, while S denotes the soft gluon interactions.}
\label{fig:fsa}
\end{figure}

The decay rate is proportional to the imaginary part
of the forward amplitude shown in Fig. \ref{fig:fsa}. 
The hard amplitude involving the final
state quark is denoted by the shaded blob which also includes the lepton loop. The
soft gluon interactions responsible for the infrared sensitivity dress the hard
amplitude, connecting to it and to the heavy quark. We use the
fact that for the soft gluon of momenta $k
\ll m_b$, we may perform the sum over cuts implicit in taking the imaginary part,
for just the hard part of the amplitude leaving the soft gluon lines uncut. 
Thus we sum over the cuts of the hard amplitude only and leave the soft 
gluon propagators uncut in the diagrams for the forward scattering amplitude. 
One can show that up to linear accuracy,
after summing over the cuts of the hard part, and integrating over its phase
space and that of the leptons, the interaction of soft gluon radiation 
with the hard part may be replaced by effective local vertices. 

\begin{figure}
\def\epsfsize#1#2{0.75#1}
\centerline{\epsfbox{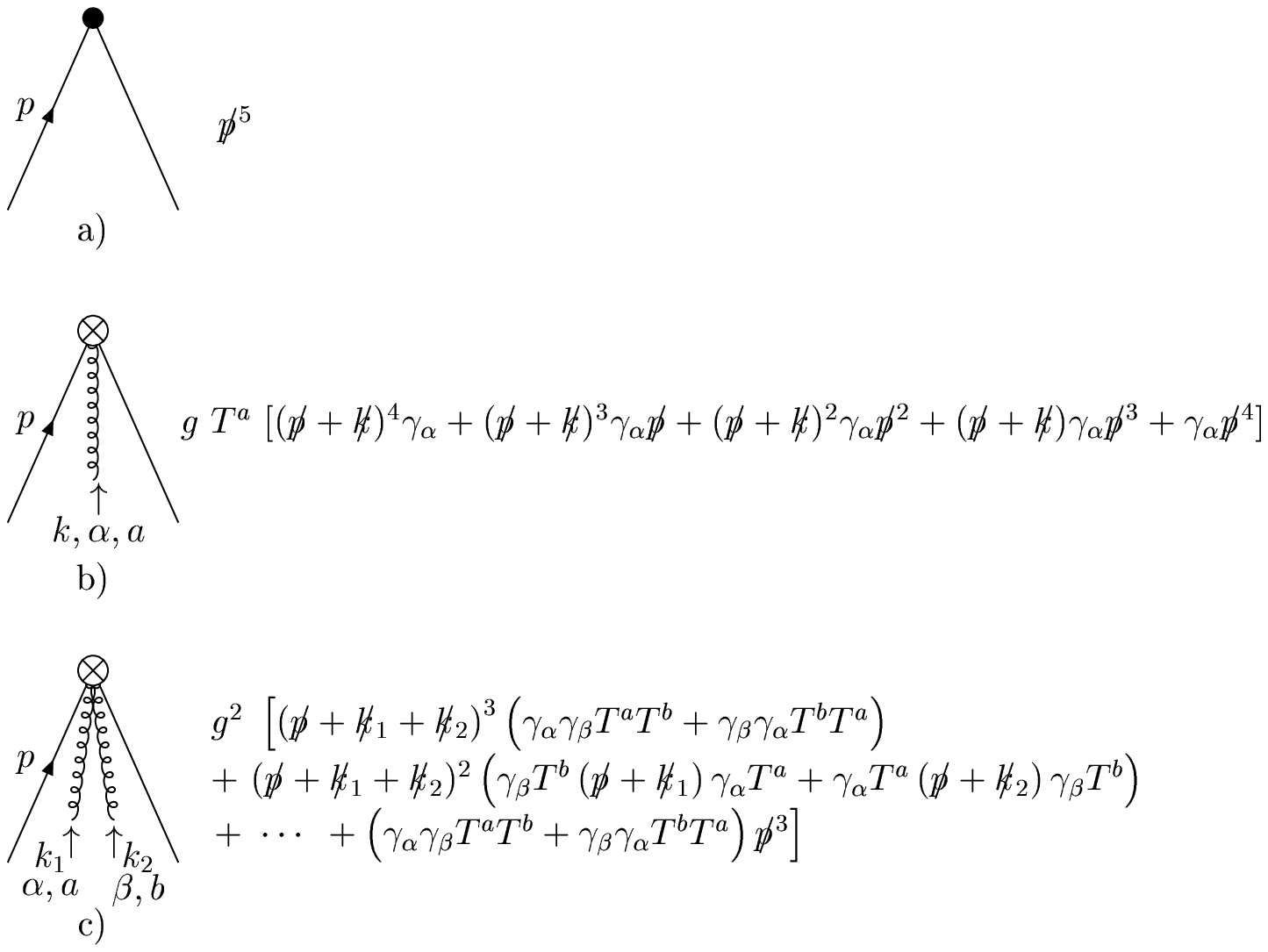}}
\bigskip

\caption{\setlength{\baselineskip}{0.30in}
a) leading effective vertex,  
b) single gluon effective vertex ,
c) double gluon effective vertex}
\label{fig:effvertex}

\end{figure}

In the parton model, the total width for the semileptonic decay is to lowest
order:
\be
\Gamma_0~=~{ G_F^2|V_{ub}|^2 \over 192 \pi^3}m_b^5
\ee
We may represent this by an effective local vertex, obtained by
preforming the phase space integration over the leptons and and the massless
quark. Such an effective vertex may be written as:
\be
C\bar{u}\ps^5 u \label{v0},~~~~~~C~=~{ G_F^2|V_{ub}|^2 \over 192 \pi^3}.
\ee
where $u$ is the heavy quark spinor.
This vertex shown on Fig. \ref{fig:effvertex}a) and will be referred to as the leading effective vertex.

Similarly, effective vertices can be derived for the
absorbtion of one and two gluons from the final state quark.
(Figure \ref{fig:effvertex}b), \ref{fig:effvertex}c)). 
These effective vertices may be obtained from the leading one by
gauge invariance, by means of the replacement $p_{\mu} \rightarrow
(p_{\mu}-gT^aA^a_{\mu})$. 
In the following we study the infrared contribution of
the effective vertex diagrams and show the cancellation of infrared sensitive
pieces linear in an infrared cutoff. 

All the infrared cancellations are shown strictly algebraically  
at the level of the corresponding  Feynman integrands. Power counting is used to
isolate and to show the cancellations of the possibly infrared sensitive
contributions upto linear accuracy. 
In this way we avoid using any
particular IR cutoff, which is crucial for the case of non-abelian theory. 


\section{Cancellation of Infrared Sensitive Terms Upto Linear Accuracy from 
the Wave Function Renormalization Diagrams} \label{sec:wave}

In this section we will show that the infrared sensitive terms upto linear
accuracy are cancelled between the wave function renormalization contributions
from diagrams with self energy insertions and parts of bremsstrahlung diagrams.  
Since the wave
function renormalization is multiplicative, all its contributions factorize from
the hard amplitude. The corresponding
bremsstrahlung diagrams that cancel the logarithmic and linear dependence on the
IR cutoff of the wave function renormalization diagrams may be obtained by
certain Ward-like identities. 
In this talk I explicitly show our method on the the one-loop order 
cancellation only. The procedure can be generalized to second loop order,
which is the actual new result.

We will denote by $-i\Sigma(p)$ the sum of all 2 point one
particle irreducible graphs. Then the full fermion propagator is:
\be
S_F^{'}(p)~=~{i \over \ps-m_b-\Sigma(p)}
\ee
Near the mass shell we have
\be
\Sigma(p)~=~\delta m\left(m_b,\Delta m \right) - (Z^{-1}-1)
\left(\ps-m_b \right) + O\left((\ps-m_b)^2 \right)
\ee
where $\Delta m$ denotes the mass counterterm:
\be
\delta m\left(m_b,\Delta m \right) \left. \right|_{\ps=m_b}~=~0.
\ee
Thus $m_b$ denotes the pole or the physical mass:
\be
\lim_{\ps \rightarrow m_b}S_F^{'}(p)~=~{i \over \ps-m_b}Z
\ee
From the LSZ reduction formula, each external line in the S-matrix element 
gets a factor $Z^{1/2}$ and we are interested in its perturbative expansion. 
Thus we expand,
\be
Z^{1/2}~=~1-{1 \over 2}\left(Z^{-1}-1 \right)+{3 \over 8}\left(Z^{-1}-1
\right)^2+..... \label{wf1}
\ee
and $ Z^{-1}-1 $ may be computed using:
\be
Z^{-1}-1~=~-{\bar{u}{\partial \over \partial p_{\nu}}\Sigma \left.
\right|_{\ps=m_b} u \over
\bar{u}\gamma^{\nu}u} \label{wf2}
\ee
In the rest frame of the heavy quark, only $\nu=0$ contributes.

\begin{figure}
%
%
%
%
%
%
%

\def\epsfsize#1#2{0.75#1}
\centerline{\epsfbox{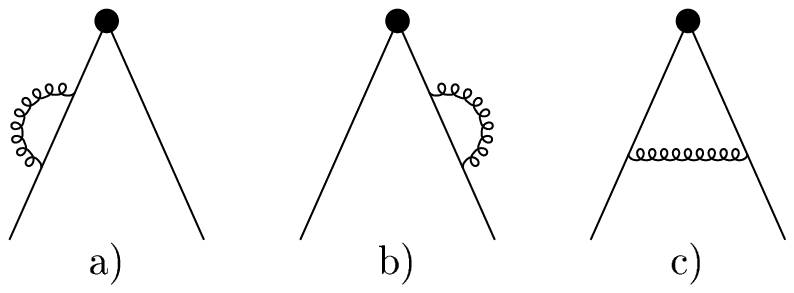}}
\bigskip

\caption{\setlength{\baselineskip}{0.30in}
Wave function renormalization and Bremsstrahlung diagrams 
to order $g^2$.} 
\label{fig:wave1}

\end{figure}

\begin{figure}
%
%
%
%
%
%
%

\def\epsfsize#1#2{0.75#1}
\centerline{\epsfbox{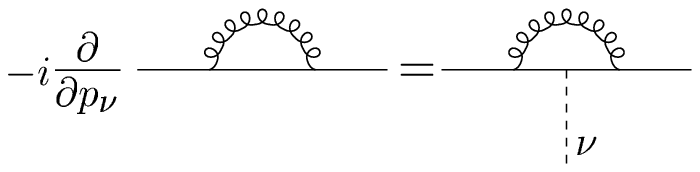}}
\bigskip

\caption{\setlength{\baselineskip}{0.30in}
One-loop order Ward-like identity}
\label{fig:ward1}

\end{figure}

Consider the one-loop order wave function renormalization and the corresponding
bremsstrahlung diagrams shown on Fig. \ref{fig:wave1}. 
The wave function renormalization parts of both
graph(a) and graph(b) are the hard part times a factor of $Z^{1/2}$. Thus we 
expand it using (\ref{wf1}) keeping only the first two terms and computing the
perturbative contribution to order $\alpha$ of $Z^{-1}-1$ from (\ref{wf2}).
Next using the identity
\be
{\partial \over \partial p^{\nu}}{i \over \ps-\ks-m_b}~=~
{i \over \ps-\ks-m_b}i\gamma_{\nu}{i \over
\ps-\ks-m_b}
\ee
we obtain the Ward-like identity  shown in Fig. \ref{fig:ward1}, 
where the dotted line denotes the insertion of a zero momentum vertex 
$\gamma_{\nu}$. 
Going to the rest frame of the heavy quark and using 
Eqs. (\ref{wf1}), (\ref{wf2}) we immediately see

\be
\left. \ref{fig:wave1}c) \right|_{wave function}= (Z^{-1}-1)_{g^2} Cm_b^5 \ub u
= -\left[ \ref{fig:wave1}a) + 
\ref{fig:wave1}b) \right].
\ee
Here, and henceforth, unless otherwise specified, the equality sign is for terms
that are logarithmically and linearly infrared sensitive. This is the desired
result, which does not involve using an explicit infrared cutoff. 

\begin{figure}
%
%
%
%
%
%
%
%
%
%
%
%
%
%
%
%
%
%
%
%
%
%
%

\def\epsfsize#1#2{0.75#1}
\centerline{\epsfbox{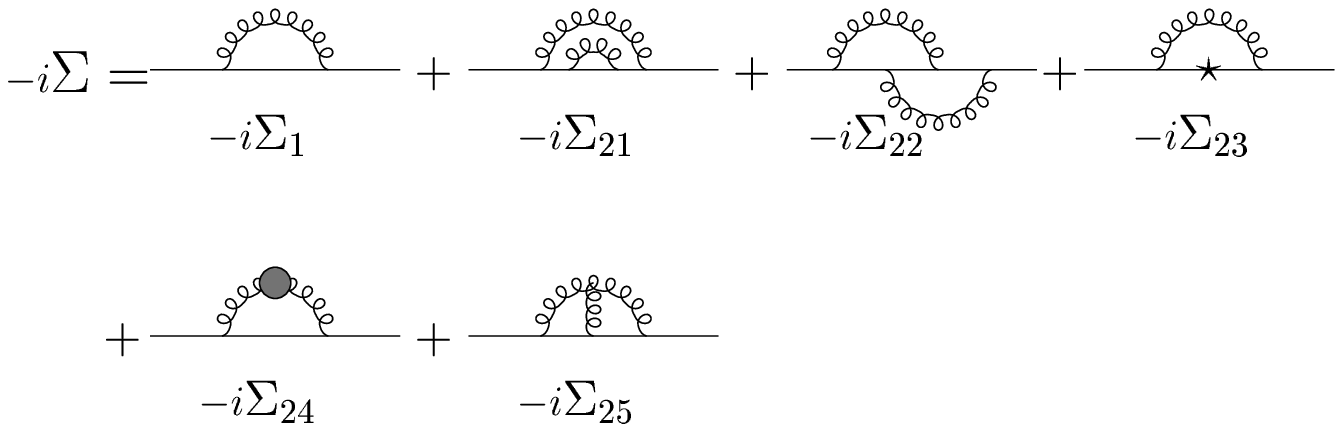}}
\bigskip

\caption{\setlength{\baselineskip}{0.30in}
Self-energy corrections of order $g^2$ and $g^4$}  
\label{fig:selfen}
\end{figure}

\begin{figure}
%
%
%
%
%
%
%
%

\def\epsfsize#1#2{0.75#1}
\centerline{\epsfbox{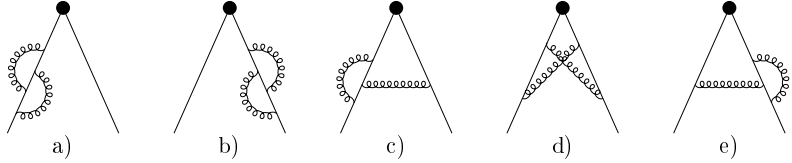}}
\bigskip

\caption{\setlength{\baselineskip}{0.30in}
The second group of 2-loop diagrams corresponding to $\Sigma_{22}$.}
\label{fig:wave22}

\end{figure}

\begin{figure}

\def\epsfsize#1#2{0.75#1}
\centerline{\epsfbox{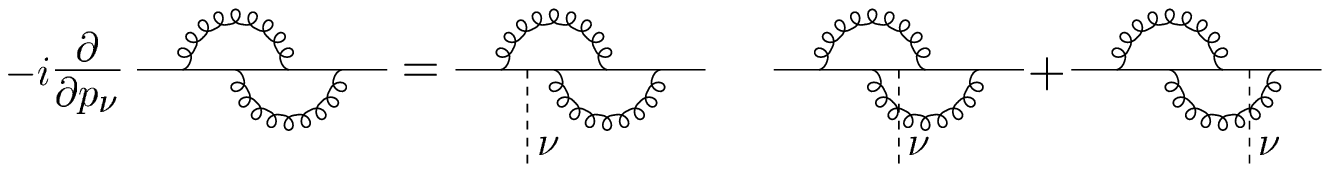}}
\bigskip

\caption{\setlength{\baselineskip}{0.30in}
Ward-like identity to the second group of 2-loop diagrams}
\label{fig:ward2}

\end{figure}

The cancellation of the wave function parts at two loops follows the same 
general method except that now we have more diagrams.
The 2-loop one particle irreducible diagrams can be divided
into  5 groups according to the 5 pieces of the order $g^4$  self-energy
corrections  (Fig. \ref{fig:selfen}).
These groups are essentially the different color groups, 
with the exception that the first and third group have the same color structure.
The cancellation of logarithmic and linear terms in the infrared cutoff takes 
place within each color group separately. As in the one-loop case, the 
cancellation is obtained by use of Ward-like identities, valid to 
linear accuracy. Figures \ref{fig:wave22} and \ref{fig:ward2} show one color  
group of diagrams and the corresponding Ward-like identity as an example.
As before, the infrared sensitive pieces are identified by power counting 
(no specific cutoff is used) and the cancellation is algebraic at the level 
of integrands.

At the two-loop level 1-particle reducible diagrams also contribute to 
the wave function renormalization. 
The contribution of the diagrams 
can be simply calculated using the perturbative expansion
of $Z^{1/2}$ (Eq. \ref{wf1}) multiplying each external line in the S-matrix
element. A straightforward calculation shows that 
the wave function renormalization pieces from the two-loop order 1-particle reducible diagrams cancel.

\section{Cancellation of the Infrared Sensitive Terms from Mass
Renormalization} \label{sec:mass}
Let us now consider the cancellation of the leading infrared sensitive piece (the
one that is linearly divergent in the infrared) arising from the diagrams
involving the mass renormalization of the heavy quark. While the wave function 
renormalization was multiplicative, and so the hard part of diagrams factorized,
the mass renormalization is additive, and the structure of hard vertex is 
important here.  
As for the wave function pieces,
I discuss the cancellation in one loop order to exemplify our method, 
and indicate how to generalize the calculation for second loop order.

As discussed in the introduction, the pole mass of the heavy quark contains a
long distance piece which is proportional linearly to an infrared cutoff,
the first contribution starting at order $\alpha$. 
By infrared power counting we can isolate the linearly
infrared sensitive pieces of the self-energy diagrams as
\begin{eqnarray}
\left.\Sigma_1 \right|_{\ps=m_b} &=& a_1 g^2 + b_1 g^2 \lambda
\label{eq:sigm}\\ \nn
\left.\Sigma_{2i} \right|_{\ps=m_b} &=& a_{2i} g^4 + b_{2i} g^4 \lambda \qquad
i=1 \ldots 5
\end{eqnarray}
It is convenient to separate out this piece and define the short distance or 
running mass to one and second loop order respectively as
\begin{eqnarray}
\bar{m} &=& m_0 + a_1g^2 \\ \nn
\bar{m} &=& m_0 + a_1 g^2 + \sum_{i=1}^{5} a_{2i} g^4.
\end{eqnarray}
We should emphasize our notation at
this point. Even though one may use a gluon mass as an infrared cutoff at 
the one loop level, this is not a gauge invariant procedure beyond it. 
Thus $\lambda$ refers to some gauge invariant cutoff, which we need not specify.
Terms that are linearly divergent in the infrared are identified by power 
counting and cancellation of such terms is shown at the level of the 
corresponding Feynman integrands. 

\begin{figure}
%
%
%
%
%
%
%
%
%
%

\def\epsfsize#1#2{0.75#1}
\centerline{\epsfbox{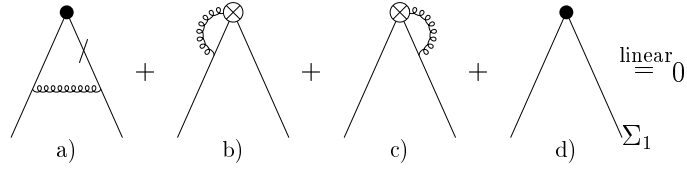}}
\bigskip

\caption{\setlength{\baselineskip}{0.30in}
Mass renormalization and the corresponding bremsstrahlung diagram to 
order $g^2$.
Figure a) depicts the mass renormalization piece of the graph. The
propagator with slash is cancelled by the expansion of the 
leading effective vertex.
The subscript $\Sigma_1$ of d) denotes the mass
shift due to the order $g^2$ self-energy correction.} 
\label{fig:mass1}
\end{figure}

The various diagrams contributing to this order are shown in 
Fig. \ref{fig:mass1}.
Fig. \ref{fig:mass1}a) is the bremsstrahlung contribution. 
To find the leading infrared sensitive contribution, we expand the leading 
effective hard vertex to first order in the offshellness,
\be
C (\ps-\ks)^5 = C (\ps-\ks-m_b+m_b)^5 =
C m_b^5+ 5 C m_b^4(\ps-\ks-m_b) + \cdots
\ee
This is an expansion of the leading effective vertex in  
$(\ps-\ks-m_b) / m_b$ and it is understood that we keep such deviations 
from the mass shell in as much as they
cancel the corresponding small denominators from the propagators. 
Then after this expansion, one of the fermion propagators in the integrand 
of the expression for Fig. \ref{fig:mass1} is cancelled (shown by a slash 
in the Figure) and we are left with an expression which resembles that for 
the self energy:
\be
\ref{fig:mass1}a)=5 C(\bar{m})^4b_1g^2\lambda \ub u
\ee
Next we have the contributions from the bremsstrahlung from the final state quark
which as discussed earlier is replaced by the effective vertices for the gluon
emission. The two diagrams with single gluon effective vertices of Figs. 
\ref{fig:mass1}b) and \ref{fig:mass1}c) give: 
\be
\ref{fig:mass1}b) + \ref{fig:mass1}c) = 2 \times C(-i5m_b^4) \bar{u} \left(-i
\Sigma_1(p) \right)u
 = -2 \times 5 C(\bar{m})^4b_1g^2\lambda \ub u.
\ee
Finally, in the lowest order diagram, Fig. \ref{fig:mass1}d), 
we must express the leading effective vertex in terms of the short distance 
mass,
\be
\ref{fig:mass1}d)~=~Cm_b^5= C(\bar{m} + b_1g^2\lambda)^5 = C \bar{m}^5 \ub u+ 
5 C (\bar{m})^4 b_1g^2\lambda \ub u.
\ee
Adding together all the order $g^2$ terms we see the cancellation of the
linearly infrared divergent pieces to this order when the decay rate is 
expressed in terms of the running mass $\bar{m}$. 

\begin{figure}

\def\epsfsize#1#2{0.75#1}
\centerline{\epsfbox{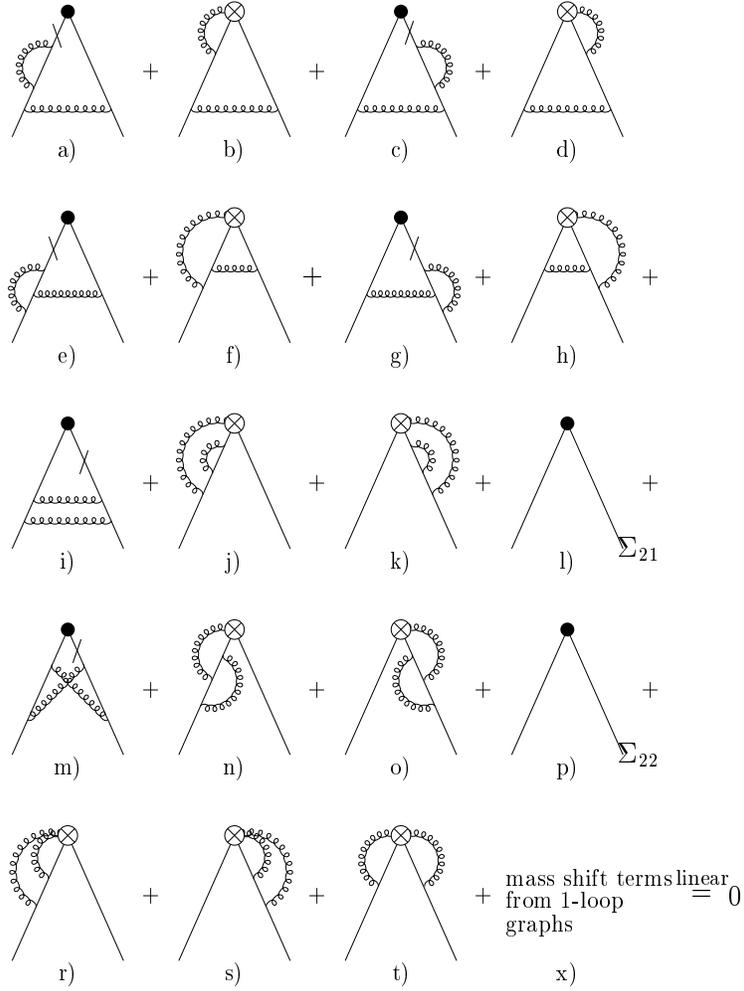}}
\bigskip

\caption{\setlength{\baselineskip}{0.30in}
Mass renormalization of order $g^4$: group-1 and group-2. 
The linear infrared sensitive mass renormalization pieces of 
the leading effective vertex graphs cancel with the 
corresponding single and double gluon vertex graphs and with $g^4$ 
order mass shift terms from the 1-loop diagrams.}
\label{fig:mass21} 
\end{figure}

\begin{figure}

\def\epsfsize#1#2{0.75#1}
\centerline{\epsfbox{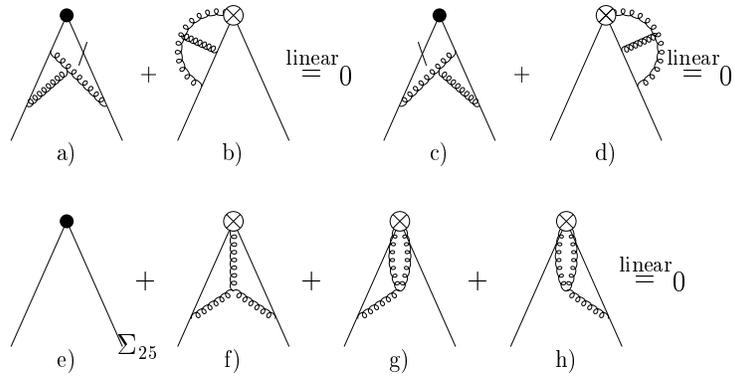}}
\bigskip

\caption{\setlength{\baselineskip}{0.30in}
Mass renormalization of order $g^4$: cancellation of linear terms 
in group-5.}
\label{fig:mass25} 
\end{figure}

As in the wave function renormalization, we group the two-loop 1 PI 
diagrams according to the five self-energy pieces shown in Fig. 
\ref{fig:selfen}. These groups are now enlarged by new diagrams with the
single and double gluon vertices 
(Fig. \ref{fig:effvertex}). Unlike in the wave function case, where each 
group of graphs cancelled separately, the cancellation of linear  
infrared sensitive pieces in the mass renormalization
involves mixing between the different groups and mass shift terms
from one loop graphs. 
To show the cancellation of linearly infrared
sensitive pieces, we follow the same procedure as in the one-loop case:
expand the effective vertices in the offshellness, keep terms that would be 
linearly proportional to an infrared cutoff, and show the cancellation 
algebraically. Two cancelling sets are shown on Figures \ref{fig:mass21},
and \ref{fig:mass25}. There are two additional sets of diagrams with abelian
type of cancellation (the cancellation proceeds in the same way as the 
interaction was abelian).
The second loop order calculation explicitely verifies that cancellation 
of linearly infrared pieces occurs if expressing the decay width in terms of 
the short distance (running) mass.

\section{Summary and Conclusions} \label{sec:summary}
We have shown by explicit perturbative calculations upto the second
loop order that when the inclusive decay width of a heavy quark is expressed in
terms of the short distance mass, there is complete accord with the operator product
expansion and the leading power corrections are of order $1/m_b^2$. 
 
As emphasized earlier, the above result
is true not just for the semileptonic decay considered in this paper but for any
inclusive decay with a single heavy particle initial state, 
like for example the radiative one. Such a cancellation  was
linked to the KLN theorem which had been suggested earlier
\cite{az1} to be the general principle behind  this for inclusive enough
observables.  

This work was supported in part by the US Department
of Energy.



%
%
%
%
%
%
%
%

%
%
%
%
%
%
%
%





\end{document}